\documentstyle[11pt,newpaspmod,twoside]{article}
\def\edcomment#1{\iffalse\marginpar{\raggedright\sl#1\/}\else\relax\fi}
\reversemarginpar

\begin{document}
\title{Star Clusters in Local Group Galaxies -- Impact of Environment on
Their Evolution and Survival}
 \author{Eva K.\ Grebel$^{1,2}$}
\affil{$^1$University of Washington, Department of Astronomy, Box 351580, 
Seattle, WA 98195, USA}
\affil{$^2$Hubble Fellow}

\begin{abstract}
Star clusters are found in $\sim$40\% of the  
Local Group galaxies.  Their properties are reviewed. 
The impact of galaxy environment on the evolution
and survival of star clusters is discussed.  Possible
evidence for cluster formation triggered by galaxy interactions, 
gradients in cluster size, metallicity and stellar content with 
galactocentric radius and variations as a function of galaxy type
are briefly summarized.
\end{abstract}


\section{Introduction}

Star clusters in Local Group galaxies comprise a wide range of roughly
coeval stellar
agglomerates ranging from globular clusters to associations, from very 
metal-poor objects to clusters with solar and higher metallicity, and
from ancient populations to embedded young clusters.  
Three basic types of such clusterings are observed: globular 
clusters, open clusters, and associations.  

\subsection{Globular Clusters}

Globular clusters are centrally concentrated, spherical systems with 
masses of $4.2 \la \log M [M_{\odot}] \la 6.6$ and tidal radii ranging 
from $\sim 10$ to $\sim 100$ pc. 
They are bound, long-lived ($\ga$10 Gyr) objects whose lifetimes 
may extend to a Hubble time or beyond, though due to various efficient
destruction mechanisms the presently observed globulars are likely
just a lower limit of the initial number of globular clusters.  Globular 
clusters have been observed in all known types of galaxies, but many of
the less massive galaxies do not contain globular clusters.

Thirteen of the 36 Local Group
galaxies have globular clusters.  One of the most massive, most luminous
($M_V$=$-10\fm55$) globular clusters is the old, metal-rich, 
elliptical globular Mayall {\sc ii} (or G1) in M31 (Rich et al.\ 1996).   
In contrast, the faintest ($M_V \sim 0\fm2$), least massive globular 
currently known is AM-4, a distant old Galactic halo globular cluster (Inman 
\& Carney 1987).  

\subsection{Open Clusters}

Open clusters have masses of $3 \la \log M [M_{\odot}] \la 5$ and radii
of 1 to 20 pc.  Most Milky Way open clusters survive for only $\sim$200 Myr 
(Janes, Tilley, \& Lyng\aa\ 1988), although there are long-lived open clusters 
with ages of several Gyr (Phelps, Janes, \& Montgomery 1994).
Loose, extended open clusters are dominant in spiral galaxies like the
Milky Way, while the Magellanic Clouds contain a large number of 
blue, compact, populous clusters. Open clusters were identified in 
$\sim 42$\% of the Local Group galaxies.  None were found in
dwarf spheroidals (dSphs), which are dominated by populations older 
than a few Gyr.  

The oldest open clusters (e.g., NGC\,6791: $\sim 8$ Gyr; Chaboyer, Green,
\& Liebert 1999) and the youngest globulars (e.g., Ter\,7: $\sim 8$ Gyr; 
Buonanno et al.\ 1998a) in the Local Group overlap in age.  The distinction 
between massive open clusters and low-mass globular clusters is somewhat 
arbitrary and largely based on age.  Bound objects that survive for 
more than 10 Gyr are generally called globulars even though they may resemble
sparse open clusters.

\subsection{Super Star Clusters}

Super star clusters are young populous clusters with masses exceeding 
several $10^4 M_{\odot}$ within a radius of 1--2 pc.  They may be 
progenitors of globular clusters.  The most massive super star clusters
are found in interacting and starburst galaxies (e.g., Schweizer \& 
Seitzer 1998, Gallagher \& Smith 1999).  Super star clusters
are often located in giant H\,{\sc ii} regions or as 
nuclear star clusters near the centers of massive galaxies.  Some 
may be progenitors of globular clusters.
Not every giant H\,{\sc ii} region harbors a super star cluster though.

Only a few, low-mass super star clusters are known in the Local Group.  
Giant H\,{\sc ii} regions such as 30\,Doradus in the Large Magellanic Cloud 
(LMC) and NGC\,3603 in the Milky Way both contain starburst clusters, which
interestingly show evidence for the formation of low-mass stars (e.g., 
Brandl et al.\ 1999) despite the presence of many very massive stars such
as O3 and main-sequence Wolf-Rayet stars.  If the R\,136 cluster in 30\,Dor 
can remain bound over a Hubble time it will evolve into a low-mass globular 
cluster.  Some of the populous LMC clusters might survive as well for a 
Hubble time (Goodwin 1997a). 

The Quintuplet and Arches clusters near the Galactic center (Figer et al.\
1999) are examples for nuclear clusters in the Local Group. 
They may have formed through colliding giant molecular clouds
and are expected to have a limited lifetime due to the strong tidal forces
of their environment.    

\subsection{Associations}

Star formation in giant molecular clouds leads usually to the formation of 
associations and/or open clusters, which appear to be the major    
contributors to a galaxy's field population.  
Associations are extended, unbound, coeval groups of stars with radii of
$\la 100$ pc.  They disperse on time scales of $\sim$100 Myr.   
Associations mark spiral arms in spiral galaxies.  They may be located 
within or at the edges of shells and supershells in spirals and irregulars.  
Often they are embedded in hierarchical structures of similar age such as 
stellar aggregates ($\sim$250 pc radius) and star complexes ($\sim$600 pc 
radius; Efremov, Ivanov, \& Nikolov 1987).  It seems reasonable to assume that 
associations were present in all types of galaxies during and after 
episodes of star formation.  

In the Local Group associations can be found in all galaxies with current 
star formation.  Both open clusters and associations generally have initial 
mass functions (IMFs) consistent with a Salpeter slope for high and 
intermediate-mass stars (Massey, Johnson, \& DeGioia-Eastwood 1995a, Massey
et al.\ 1995b).   

\section{Star Clusters in the Local Group}

The global properties of the star clusters in Local Group galaxies 
are summarized in Table 1.  Within a zero-velocity surface of 1.2 Mpc
(Courteau \& van den Bergh) around the Local Group barycenter 36 galaxies
are currently known.  In less than half of these globular or open 
clusters were detected. 
While past surveys established that the majority of the least 
massive galaxies, the dSphs, do not contain star clusters, the census 
for more massive galaxies is still incomplete.

\begin{table}
\caption{Global Properties of Star Clusters in the Local Group}\medskip
\tiny
\begin{tabular}{llcrcrcrcc}
Galaxy    & Type       & $M_V$ & $N_{GC}$    & $S_N$ & Ages  & [Fe/H]         & $N_{OC}$ & Ages    & [Fe/H]         \\
          &            & [mag] &             &       & [Gyr] & [dex]          &           & [Gyr]   & [dex]          \\
\\
\tableline
\\
M31       & Sb\,{\sc i-ii}     & --21.2 & $\sim$600 & 2: & $\le 15$ & --2.5 -- $0.4^a$ & many & .004--? & ? \\ 
Galaxy    & S(B)bc\,{\sc i-ii} & --20.9 & $\sim$160 & 0.7 & 8--15 & --2.5 -- 0.4 & $>$1000 & .001--9 & --1.0 -- 0.4 \\
M33       & Sc\,{\sc ii-iii}   & --18.9 & $\ga$54$^b$ & 1.5: & $\le 12$ & --3.0 -- --$0.8^c$ & $>600^c$ & .004--? & ? \\
LMC       & Ir\,{\sc iii-iv}   & --18.5 & $\sim$13  & 0.5 & 9--15 & --2.3 -- --1.2$^d$ & $\ga$4000 & .001--4$^e$ & --1.4 -- --$0.1^f$ \\  
SMC       & Ir\,{\sc iv/iv-v}  & --17.1 & 1         & 0.1 & 12    & --1.4             & $\ga$2000 & .001--10 & --1.4 -- --0.5 \\
IC\,10    & Ir\,{\sc iv}:      & --16.3 & 0         & 0 & ---   &  ---                & several & ?  & ?  \\
NGC\,6822 & Ir\,{\sc iv-v}     & --16.0 & 1         & 0.4 & $\sim$11 & --2.0$^g$ & $\sim$30$^h$ & 2 ({\sc vii})$^g$ & --1.0 ({\sc vii})$^g$ \\ 
IC\,1613  & Ir\,{\sc v}        & --15.3 & 0         & 0 & ---   & ---                 & $>6^i$ & .01--? & ? \\
WLM       & Ir\,{\sc iv-v}     & --14.4 & 1         & 1.7 & 15    & --$1.5^j$           &   ?    & ? & ?  \\
NGC\,205 & Sph        & --16.4 & $\ga$14$^k$ & $\ga$3.9 & ? & --1.9 -- --$1.3^l$ & $\ge$2$^m$ & $\ge$.05$^m$ & ? \\ 
NGC\,185 & Sph        & --15.6 & $\ga$8$^k$ & $\ga$4.6 & ? & --2.5 -- --1.2$^l$ & some$^k$ & ?  & ?  \\
NGC\,147 & Sph        & --15.1 & $\ge$4$^n$ & $\ga$3.6 & ? & --2.5 -- --$1.9^l$ & ?  & ?  &  ? \\
Sgr       & dSph(t)    & --13.8: & $\ge$4 & $\ge$12.1: & 8--15 & --2.0 -- --$0.4^o$ & 0 & ---  &  --- \\
For       & dSph      & --13.1 & 5 & 22.8 & 12--15 & --2.2 -- --$1.8^p$ & 0 & --- & --- \\ 
And\,I    & dSph      & --11.8 & 1 & 20.9 & old    & --$1.4^q$ & 0 & --- & --- \\
\\
\tableline
\tableline\\
\end{tabular}
{\bf Notes:}\, Galaxy types (Col.\ 2) and $M_V$ (Col.\ 3)
were taken from Courteau \& van den Bergh (1999). $N_{GC}$ 
and $N_{OC}$ (Cols.\ 4 \& 7) denote the number of globular clusters 
and open clusters, respectively.  All quoted ages assume an oldest 
age of 15 Gyr. $S_N$ is the specific globular cluster frequency.\,
{\bf References:}\, $^a$Barmby et al.\ 2000, astro-ph/9911152; 
$^b$Christian \& Schommer (1988, AJ, 95, 704); Mochejska et al.\ 1998, 
    AcA, 48, 455;
$^c$Chandar et al.\ 1999, ApJ, 517, 668; 
$^d$Olsen et al.\ 1998, MNRAS, 300, 665; $^e$Sarajedini 1998, AJ, 116, 738; 
$^f$Jasniewicz \& Th\'evenin 1994, A\&A, 282, 717;
$^g$Cohen \& Blakeslee (1998, AJ, 115, 2356); $^h$Hodge (1977, ApJS, 33, 69); 
$^i$Wyder et al.\ (2000, 33rd ESLAB Symp., in press); 
$^j$Hodge et al.\ (1999, ApJ, 521, 577);
$^k$Geisler et al.\ (1999, IAU Symp.\ 192, p.\ 231);
$^l$Da Costa \& Mould (1988, ApJ, 334, 159);
$^m$Cappellari et al.\ (1999, ApJ, 515, L17);
$^n$Hodge (1976, AJ, 81, 25).
$^o$Da Costa \& Armandroff (1995, AJ, 109, 2533);
$^p$Buonanno et al.\ (1998, ApJ, 501, L33); 
$^q$Grebel et al.\ (2000a, in prep.).
\normalsize
\end{table}

\section{Environmental Effects and Interactions Between Galaxies}

Generally, the number
of globular clusters is larger in the more massive galaxies, while
the few globular clusters in faint dSphs lead to high specific
frequencies $S_N$ (i.e., the number of globulars, $N_{GC}$, normalized by
parent galaxy luminosity; $S_N = N_{GC}\cdot
10^{0.4(M_V+15)}$, Harris \& van den Bergh 1981).  

\subsection{Tidal Stripping}

The orbital decay times of globular clusters in dSph galaxies are of
the order of only a few Gyr (Hernandez \& Gilmore 1998).
While in Sagittarius and Fornax one of the globulars 
lies near the projected galaxy center, both dSphs show spatially 
extended globular cluster systems.  This as well as the 
puzzling present-day lack of gas suggests that they underwent 
significant mass loss (Oh, Lin, \& Richer 2000).  The detection of 
extratidal stars suggests that tidal stripping may have reduced some dSphs 
to as little as 1\% of their original mass (Majewski et al.\ 2000).  
This might explain the inflated $S_N$ measured today.

The dSph And\,I is the least massive galaxy in which a globular cluster
has been detected to-date (Grebel, Dolphin, \& Guhathakurta 2000a).  Its faint,
sparse globular resembles Galactic outer halo globulars and is located just 
beyond the galaxy's core radius.  And\,I is another good 
candidate for tidal stripping due to its close proximity to M31 (45--85 kpc, 
Da Costa et al.\ 1996).  

Conversely, some globular clusters may have been stripped from dwarf 
galaxies by more massive galaxies, or may be the cores of accreted 
nucleated dwarf galaxies (e.g., Bassino, Muzzio, \& Rabolli 1994). 
Its abundance spread and possible age range suggest that the Milky Way's
most massive globular, $\omega$\,Cen, may be the result of such an
evolution (e.g., Hughes \& Wallerstein 2000). 

\subsection{Cluster Formation Triggered By Galaxy Interactions?}

The Magellanic Clouds interact with each other and with the Milky 
Way, and stand out through their excess in young populous
clusters.  No such large numbers are observed in the other, less
massive Local Group irregulars, none of which are close to a massive 
spiral.  The field star formation history of the Magellanic Clouds is
fairly continuous (Holtzman et al.\ 1999), but the LMC shows a pronounced
peak in cluster formation 1--2 Gyr ago, which coincides with the second 
last close encounters with the Milky Way (Girardi et al.\ 1995).  Curiously 
the Small Magellanic Cloud (SMC) does not show a corresponding peak at 
this age.   However, the cluster age distributions of LMC and SMC both peak
at 100--200 Myr (Grebel et al.\ 2000b), which coincides with the last
perigalacticon and the last close encounter between the Magellanic Clouds
(Gardiner, Sawa, \& Fujimoto 1994).  This appears to suggest that the recent 
increases in cluster formation may have been partially 
interaction-triggered.

\section{Cluster Destruction Through Intragalactic Environment}

Gnedin \& Ostriker's (1997) ``vital diagrams'' summarize the dominant effects 
of globular cluster destruction (see also Gerhard, these proceedings) 
in the Milky Way:  relaxation, tidal shocks 
through disk and bulge passages, and dynamical friction as a function of
globular cluster mass and half-mass radius.  At small
Galactocentric distances only fairly massive, compact globular clusters
can survive for more than a Hubble time.  This correlates well with
the observed increase in globular cluster half-light radii with Galactocentric
distance (van den Bergh 1994a), while the lack of compact clusters at large
distances remains puzzling.  Goodwin (1997b) suggests that for a given
Galactocentric radius, IMF, and star formation efficiency (SFE), the survival 
of a cluster depends on its central density at the time of formation (see
McLaughlin's contribution for more on SFEs).

The short (0.2 Gyr) lifetime of Milky Way open clusters is believed to be 
partially due to interactions with giant molecular clouds, which destroy 
open clusters 
within the solar radius (Wielen 1991).  In the Magellanic Clouds, a less
violent dynamical environment, open cluster life expectancies are much longer
(median age 0.9 Gyr (SMC) and 1.1 Gyr (LMC); Hodge 1988).   

\section{Cluster Properties as Function of Environment}

\subsection{The Oldest Star Clusters in the Local Group}

While absolute and relative age determinations for globular clusters 
are plagued by a number of uncertainties (Stetson, VandenBerg, \& Bolte 
1996; Sarajedini, Chaboyer, \& Demarque 1997), there is growing consensus that
Galactic globular clusters show age differences $\ge$5 Gyr.  Age values 
listed in Table 1 refer to an arbitrarily chosen oldest age of 15 Gyr.
The oldest globular clusters in the Milky Way, the LMC (Olsen et al.\ 1998; 
Johnson et al.\ 2000), Sagittarius (Montegriffo et al.\ 1998), and 
Fornax (Buonanno et al.\ 1998b) formed at the same time, while cluster
formation began $\sim$3 Gyr later in the SMC (Mighell, Sarajedini, \&
French 1998) and M33 (Sarajedini et al.\ 1998).  These differences are
not obviously correlated with galaxy mass, type, or location.   

\subsection{Radial Abundance Gradients}

The globular cluster systems of Local Group spirals show an overall
radial abundance gradient in the sense that the central (bulge) regions
contain a range of abundances including very metal-rich clusters (Barbuy,
Bica, \& Ortolani 1998), while
the mean metallicities decrease with increasing galactocentric distance.
The LMC appears to show similar behavior (Da Costa 2000).  

\subsection{Cluster Shapes and Sizes}

Due to the weaker tidal field of the LMC, its clusters can retain higher
ellipticities longer (Goodwin 1997c).  At a given galactocentric
distance the LMC globular clusters tend to be larger than Galactic 
globulars, and in both Milky Way and LMC half-light
radii ($r_h$) increase with galactocentric distance (van den Bergh 2000 
and Sect.\ 4).  Galactic globulars on nearly circular orbits are 
systematically larger (van den Bergh 1994a), while
clusters on retrograde orbits are
smaller than other globulars, indicative of preferred tidal
stripping (van den Bergh 1994b).  One could 
speculate that the small $r_h$ values of Fornax's globulars indicate that 
this dSph was initially much more massive.  Van den Bergh (1994a) interprets 
the small $r_h$ values as evidence against accretion of globular clusters
from disintegrating dSphs by the Galactic halo.  In contrast, the large 
$r_h$ of And\,I's globular cluster ($\sim$10 pc) is comparable to that of 
Galactic outer halo globular clusters (Grebel et al.\ 2000a).  
 
\subsection{Stellar Content}

The horizontal branch (HB) morphologies of inner LMC globular clusters resemble 
those of inner Galactic halo globulars, while outer LMC globulars show the
same second-parameter effects as their Galactic counterparts (Johnson et al.\
2000).  Two of the old Fornax globular clusters show clear second-parameter 
HB variations (Smith et al.\ 1996, Buonanno et al. 1998b).  This shows not
only that second-parameter globulars in dwarfs can be as old as the oldest
Galactic halo globulars, but also that accretion of globular clusters from 
disintegrating dwarfs (a mechanism proposed to explain the metal-poor, red HB
clusters in the outer Galactic halo) would contribute both types of globulars
to the Milky Way halo.

For a discussion of the metallicity dependence of the blue-to-red supergiant 
ratio in young Galactic and Magellanic clusters we refer to Langer \& Maeder
(1995).  A possible metallicity dependence of Be star fractions and rotation
in young clusters in these galaxies is discussed in Maeder, Grebel, \& 
Mermilliod (1999). 

\section{Summary}

In about 40\% of the Local Group galaxies star clusters have been detected
so far, but the census is still incomplete.  Several (but not all)
Local Group galaxies
seem to share a common epoch of the earliest globular cluster formation.
The most massive galaxies show a tendency for rapid 
enrichment in their oldest cluster population near their centers as compared 
to clusters at larger galactocentric radii.  The galactocentric dependence
of cluster sizes and HB morphology may be intrinsic to the globular cluster
formation process rather than to the accretion of dwarf galaxies.
The observed properties of
the globular clusters in the least massive dwarfs suggest that their parent
galaxies may originally have been substantially more massive.  
Cluster destruction mechanisms and time scales are a function of
galaxy environment and galaxy mass.   
The most recent enhancement of star cluster formation in the Magellanic 
Clouds may have been triggered by their close encounter with each other and 
the Milky Way.

\acknowledgments
It is a pleasure to acknowledge support by NASA through grant
HF-01108.01-98A from
STScI, which is operated by AURA, Inc., under NASA contract NAS5-26555.

\end{document}